\documentclass[runningheads]{svmult}

\usepackage{makeidx}   
\usepackage{graphicx}  
\usepackage{subeqnar}  
\usepackage{multicol}  
\usepackage{physprbb}  
\makeindex             

\newcommand{\greeksym}[1]{{\usefont{U}{psy}{m}{n}#1}}
\newcommand{\umu}{\mbox{\greeksym{m}}}

\def\mathnew{\mathsurround=0pt}
\def\simov#1#2{\lower .5pt\vbox{\baselineskip0pt \lineskip-.5pt
        \ialign{$\mathnew#1\hfil##\hfil$\crcr#2\crcr\sim\crcr}}}
\def\lesssim{\mathrel{\mathpalette\simov <}}

\begin{document}
\title*{Gamma-Ray Bursts as a Probe of Cosmology}
\toctitle{Gamma-Ray Bursts as a Probe of Cosmology}
\titlerunning{Gamma-Ray Bursts as a Probe of Cosmology}
\author{Donald Q. Lamb\inst{1}
\and Daniel E. Reichart\inst{2}
}
\authorrunning{Donald Q. Lamb and Daniel E. Reichart}
\institute{Department of Astronomy \& Astrophysics, University of
Chicago, 5640 South Ellis Avenue, Chicago, IL 60637
\and Department of Astronomy, California Institute of Technology,
Mail Code 105-24, 1201 East California Boulevard, Pasadena, CA 91125}

\maketitle              

\begin{abstract}
We show that, if the long GRBs are produced by the collapse of massive
stars, GRBs and their afterglows may provide a powerful probe of
cosmology and the early universe.
\end{abstract}

\section{Introduction} 

There is increasingly strong evidence that gamma-ray bursts (GRBs) are
associated with star-forming galaxies [1,2,3,4] and occur near or in
the star-forming regions of these galaxies [2,3,4,5,6].  These
associations provide indirect evidence that at least the long GRBs
detected by BeppoSAX are a result of the collapse of massive stars. 
The discovery of what appear to be supernova components in the
afterglows of GRBs 970228 [7,8] and 980326 [9] provides tantalizing
direct evidence that at least some GRBs are related to the deaths of
massive stars, as predicted by the widely-discussed collapsar model of
GRBs [10,11,12,13,14].  If GRBs are indeed related to the collapse
of massive stars, one expects the GRB rate to be approximately
proportional to the star-formation rate (SFR).

\section{Detectability of GRBs and Their Afterglows}

\begin{figure}
\begin{minipage}[t]{2.35truein}
\mbox{}\\
\includegraphics[width=2.35truein,clip=]{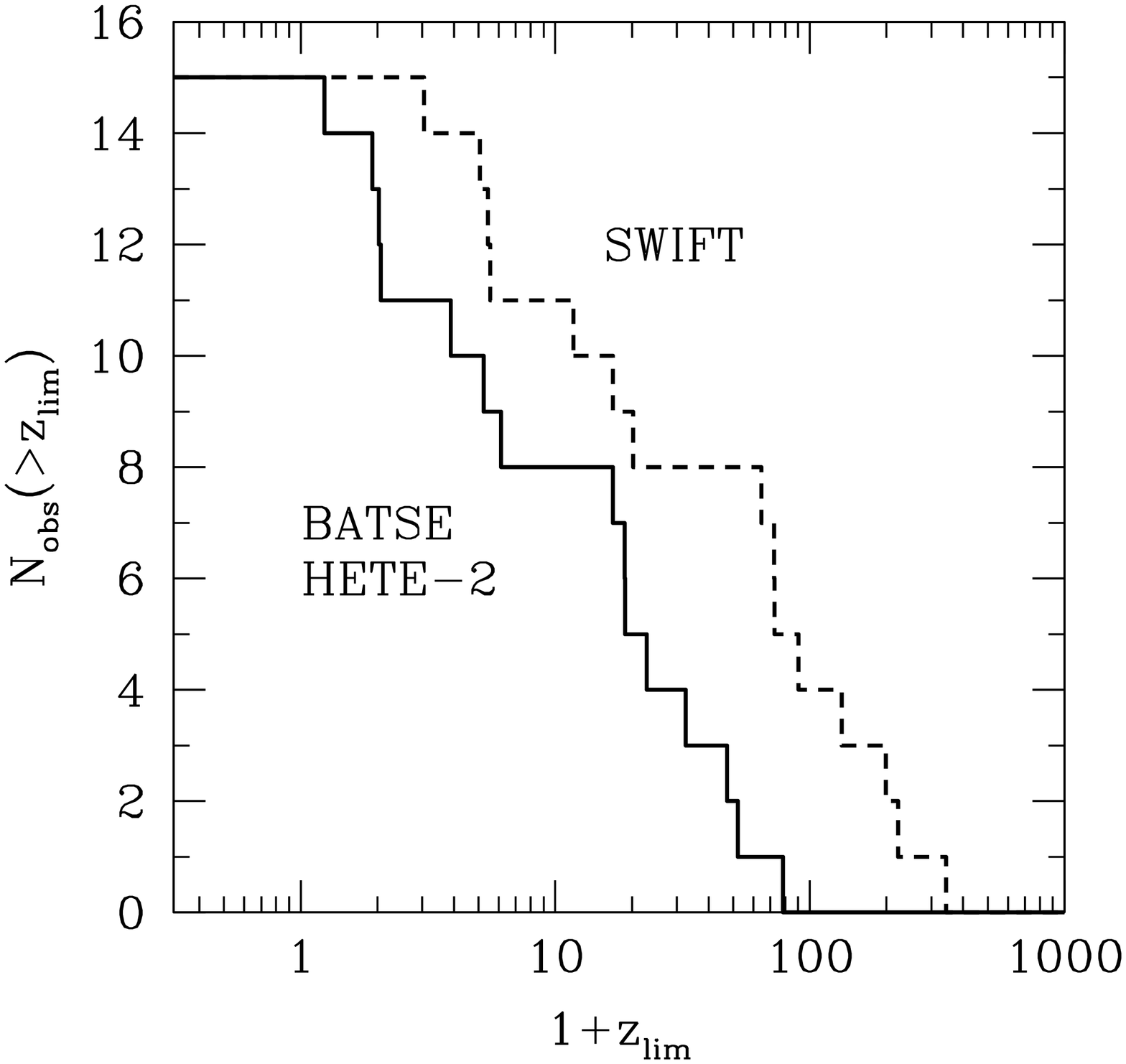}
\caption{Cumulative distributions of the limiting redshifts at which
the 15 GRBs with well-determined redshifts and published peak photon
number fluxes would be detectable by BATSE and HETE-2, and by {\it
Swift}.}
\end{minipage}
\hfill
\begin{minipage}[t]{2.35truein}
\mbox{}\\
\includegraphics[width=2.435truein,clip=]{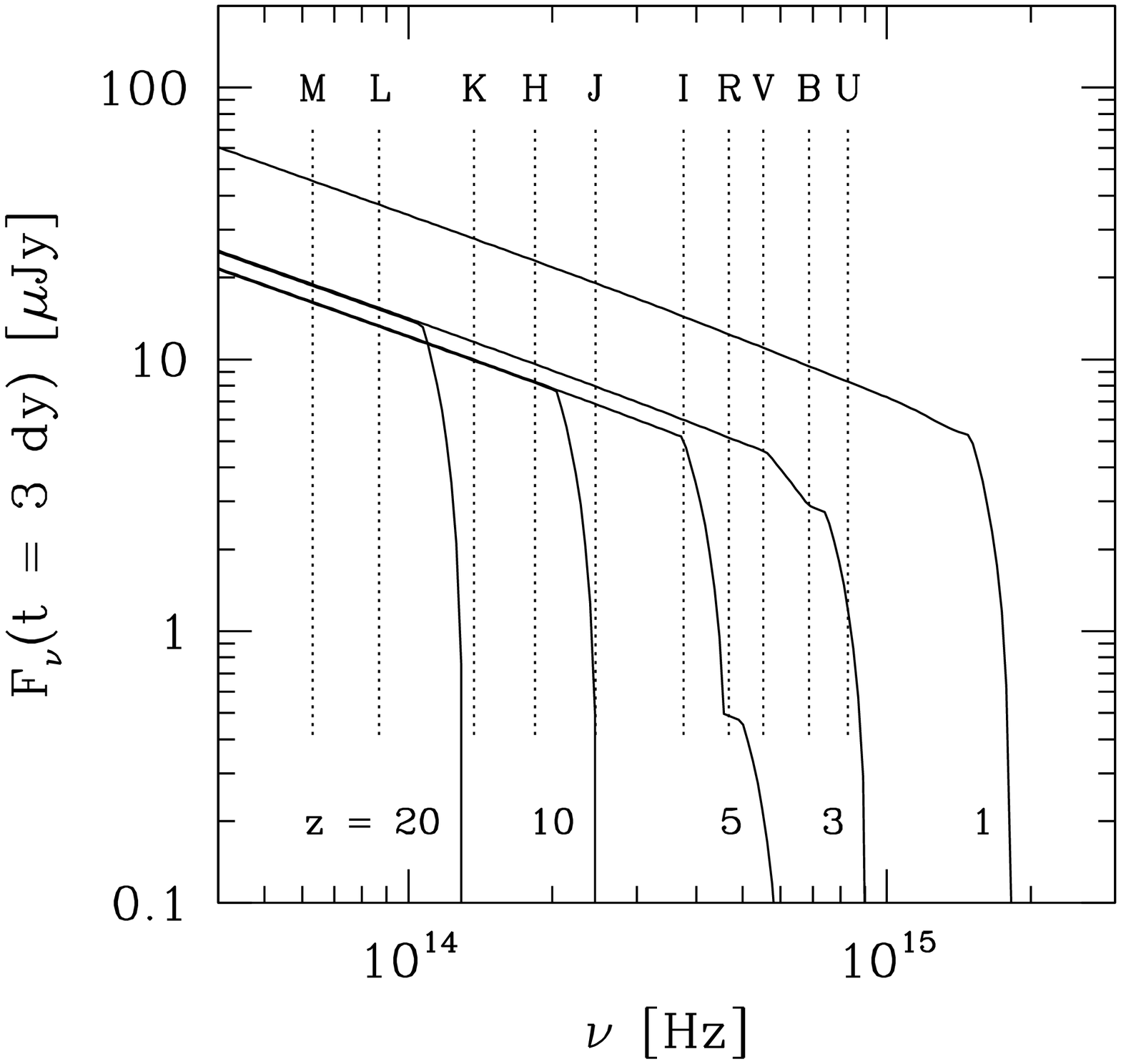}
\caption{The best-fit spectral flux distribution of the early afterglow
of GRB 000131, as observed one day after the burst, after transforming
it to various redshifts, and extinguishing it with a model of the
Ly$\alpha$ forest.}
\end{minipage}
\end{figure}

We have calculated the limiting redshifts detectable by BATSE and
HETE-2, and by {\it Swift}, for the sixteen GRBs with well-established
redshifts and published peak photon number fluxes.  In doing so, we
have used the peak photon number fluxes given in Table 1 of [15], taken
a detection threshold of 0.2 ph s$^{-1}$ for BATSE and HETE-2 and 0.04 
ph s$^{-1}$ for {\it Swift}, and set $H_0 =  65$ km s$^{-1}$
Mpc$^{-1}$, $\Omega_m = 0.3$, and $\Omega_{\Lambda} = 0.7$ (other
cosmologies give similar results).  Figure 1 displays the results. 
This figure shows that BATSE and HETE-2 would be able to detect half of
these GRBs out to a redshift $z = 20$ and 20\% of them out to a
redshift $z =50$. {\it Swift} would be able to detect half of them out
to redshifts $z = 70$, and 20\% of them out to a redshift $z = 200$,
although it is unlikely that GRBs occur at such extreme redshifts. 
Consequently, if GRBs occur at very high ($z > 5)$ redshifts (VHRs),
BATSE has probably already detected GRBs at these redshifts, and HETE-2
and {\it Swift} should detect them as well.

The soft X-ray, optical and infrared afterglows of GRBs are also
detectable out to VHRs.  The effects of distance and redshift tend to
reduce the spectral flux in GRB afterglows in a given frequency band,
but time dilation tends to increase it at a fixed time of observation
after the GRB, since afterglow intensities tend to decrease with time. 
These effects combine to produce little or no decrease in the spectral
energy flux $F_{\nu}$ of GRB afterglows in a given frequency band and
at a fixed time of observation after the GRB with increasing redshift:
\begin{equation}
F_{\nu}(\nu,t) = \frac{L_{\nu}(\nu,t)}{4\pi D^2(z) (1+z)^{1-a+b}},
\end{equation}
where $L_\nu \propto \nu^at^b$ is the intrinsic spectral luminosity of
the GRB afterglow, which we assume applies even at early times, and
$D(z)$ is the comoving distance to the burst.  Many afterglows fade
like $b \approx -4/3$, which implies that $F_{\nu}(\nu,t) \propto
D(z)^{-2} (1+z)^{-5/9}$ in the simplest afterglow model, where $a =
2b/3$ [16].  In addition, $D(z)$ increases very slowly with redshift at
redshifts greater than a few.  Consequently, there is little or no
decrease in the spectral flux of GRB afterglows with increasing
redshift beyond $z \approx 3$.

\begin{figure}
\begin{minipage}[t]{2.5truein}
\mbox{}\\
\includegraphics[width=2.5truein,clip=]{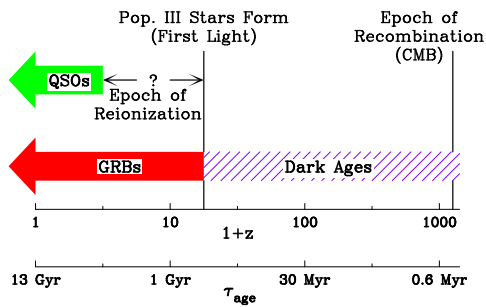}
\end{minipage}
\hfill
\begin{minipage}[t]{2.truein}
\mbox{}\\
\caption{Cosmological context of VHR GRBs.  Shown are the epochs of
recombination, first light, and re-ionization.  Also shown are the
ranges of redshifts corresponding to the ``dark ages,'' and probed
by QSOs and GRBs. }
\end{minipage}
\end{figure}

In fact, in the simplest afterglow model where $a = 2b/3$, if the
afterglow declines more rapidly than $b \approx 1.7$, the spectral flux
actually {\it increases} as one moves the burst to higher redshifts! 
An example of this is the afterglow of GRB 000131.  Its peak flux
$F_{\rm peak}$ was in the top 5\% of all BATSE bursts and the break
energy $E_{\rm break}$ in its spectrum was 164 keV, yet it occurred at
a redshift $z = 4.50$.  We have calculated the best-fit spectral flux
distribution of the afterglow of GRB 000131 from [17], as observed three
days after the burst, transformed to various redshifts.  The
transformation involves (1) dimming the afterglow, (2) redshifting its
spectrum, (3) time dilating its light curve, and (4) extinguishing the
spectrum using a model of the Ly$\alpha$ forest (for details, see
[15]).  Finally, we have convolved the transformed spectra with a top
hat smearing function of width $\Delta \nu = 0.2\nu$.  This models
these spectra as they would be sampled photometrically, as opposed to
spectroscopically; i.e., this transforms the model spectra into model
spectral flux distributions.

Figure 2 shows the resulting spectral flux distribution.  The spectral
flux distribution of the afterglow is cut off by the Ly$\alpha$ forest
at progressively lower frequencies as one moves out in redshift.  Thus 
high redshift afterglows are characterized by an optical ``dropout''
[4], and VHR afterglows by a near infrared ``dropout.''  We conclude
that, if GRBs occur at very high redshifts, both they and their
afterglows can be easily detected.

\section{GRBs as a Probe of Cosmology and the Early Universe}

Theoretical calculations show that the birth rate of Pop III stars
produces a peak in the SFR in the universe at redshifts $16 \lesssim z
\lesssim 20$, while the birth rate of Pop II stars produces a much
larger and broader peak at redshifts $2 \lesssim z \lesssim 10$
[18,19,20].  Therefore one expects GRBs to occur out to at least $z
\approx 10$ and possibly $z \approx 15-20$, redshifts that are far
larger than those expected for the most distant quasars.  

Figure 3 places GRBs in a cosmological context.  At recombination,
which occurs at redshift $z = 1100$, the universe becomes transparent. 
The cosmic background radiation originates at this redshift.  Shortly
afterwards, the temperature of the cosmic background radiation falls
below 3000 K and the universe enters the ``dark ages'' during which
there is no visible light in the universe.  ``First light,'' which
occurs at $z \approx 20$, corresponds to the epoch when the first stars
form.   Ultraviolet radiation from these first stars and/or from the
first active galactic nuclei re-ionizes the universe.  Afterward, the
universe is transparent in the ultraviolet.

\begin{figure}[t]
\begin{minipage}[t]{2.90truein}
\mbox{}\\
\includegraphics[angle=-90,width=2.90truein,clip=]{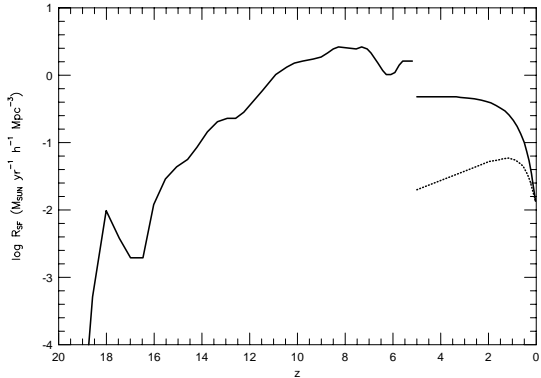}
\end{minipage}
\hfill
\begin{minipage}[t]{1.67truein}
\mbox{}\\
\caption{The cosmic SFR $R_{SF}$ as a function of redshift $z$.  The
solid curve at $z < 5$ is the SFR derived by [25]; the solid curve at
$z \ge 5$ is the  SFR calculated by [18] (the dip in this curve at $z
\approx 6$ is an artifact of their numerical simulation).  The dotted
curve is the SFR derived by [24].} 
\end{minipage}
\vskip -0.2truein
\end{figure}

QSOs are currently the most powerful probes of the high redshift
universe.  GRBs have several advantages relative to QSOs as probes of
cosmology.  First, GRBs are expected to occur out to $z \approx 20$,
whereas QSOs occur out to only $z \approx 5$.  
Second, very high redshift GRB afterglows can be 100 - 1000 times
brighter at early times than are high redshift QSOs.  This makes
possible very sensitive high dispersion spectroscopy of the metal
absorption lines and the Lyman $\alpha$ forest in the spectrum of the
afterglows. Third, no ``proximity effect'' on intergalactic distances
scales is expected for GRBs and their afterglows, in contrast to QSOs. 
Thus GRBs may be relatively ``clean'' probes of the intergalactic
medium, the Lyman $\alpha$ forest, and damped Lyman $\alpha$ clouds,
even in the vicinity of the GRBs.

The important cosmological questions that observations of GRBs
and their afterglows may be able to address include the following:  
\medskip

\noindent
$\bullet$ Information about the epoch of ``first light'' and the
earliest generations of stars from merely the detection of GRBs at very
high redshifts;
\medskip

\noindent
$\bullet$ Information about the growth of metallicity in the 
universe in the star-forming entities in which the bursts occur, in
damped Lyman $\alpha$ clouds, and in the Lyman $\alpha$ forest from
observations of the metal absorption line systems in the spectra of
their afterglows;
\medskip

\noindent
$\bullet$ Information about the large-scale structure of the universe
at VHRs from the clustering of the Lyman $\alpha$ forest lines and the
metal absorption-line systems in the spectra of their afterglows; and
\medskip

\noindent
$\bullet$ Information about the epoch of re-ionization from the
depth of the Lyman $\alpha$ break in the spectra of their afterglows.
\medskip

\noindent
Below we consider the first of these questions: the epoch of ``first
light'' and the earliest generations of stars.

\section{GRBs as a Probe of Star Formation}

Observational estimates [21,22,23,24] indicate that the SFR in the
universe was about 15 times larger at a redshift $z \approx 1$ than it
is today.  The data at higher redshifts from the Hubble Deep Field
(HDF) in the north suggests a peak in the SFR at $z \approx 1-2$ [24],
but the actual situation is highly uncertain.   

In Figure 4, we have plotted the SFR versus redshift from a
phenomenological fit [25] to the SFR derived from submillimeter,
infrared, and UV data at redshifts $z < 5$, and from a numerical
simulation by [18] at redshifts $z \geq 5$.  The simulations done by
[18] indicate that the SFR increases with increasing redshift until $z
\approx 10$, at which point it levels off.  The smaller peak in the SFR
at $z \approx 18$ corresponds to the formation of Population III stars,
brought on by cooling by molecular hydrogen.  Since GRBs are detectable
at these VHRs and their redshifts may be measurable from the
absorption-line systems and the Ly$\alpha$ break in the afterglows
[4], if the GRB rate is proportional to the SFR, then GRBs could
provide unique information about the star-formation history of the VHR
universe.

\begin{figure}[t]
\includegraphics[width=4.80truein,clip=]{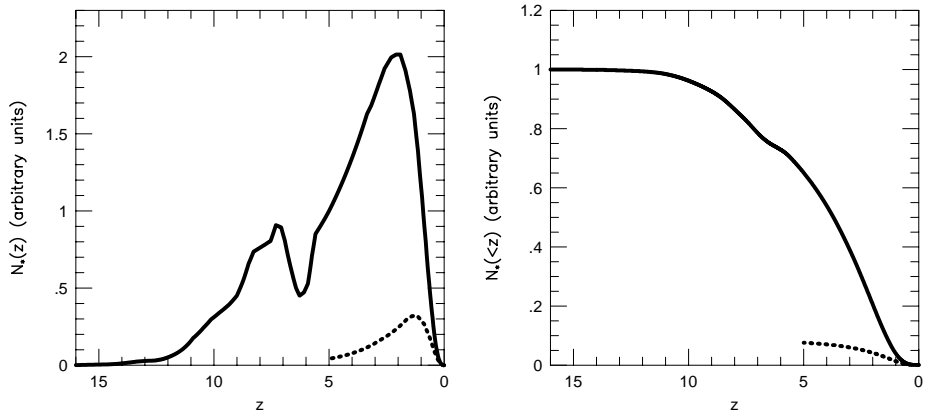}
\caption{Left panel:  The number $N_*$ of stars expected as a function
of redshift $z$ (i.e., the SFR from Figure 4, weighted
by the differential comoving volume, and time-dilated) assuming that
$\Omega_M = 0.3$ and $\Omega_\Lambda = 0.7$.  Right panel:  The
cumulative distribution of the number $N_*$ of stars expected as a
function of redshift $z$.  Note that $\approx 40\%$ of all stars have
redshifts $z > 5$.  The solid and dashed curves in both panels have the
same meanings as in Figure 4.}
\end{figure}

We have calculated the expected number $N_*$ of stars as a function of
$z$ assuming (1) that the GRB rate is proportional to the
SFR\footnote{This may underestimate the GRB rate at VHRs since it is
generally thought that the initial mass function will be tilted toward
a greater fraction of massive stars at VHRs because of less efficient
cooling due to the lower metallicity of the universe at these early
times.}, and (2) that the SFR is that given in Figure 4 (see [15] for
details).  The left panel of Figure 5 shows our results for $N_*(z)$
for an assumed cosmology $\Omega_M = 0.3$ and $\Omega_\Lambda = 0.7$
(other cosmologies give similar results).  The solid curve corresponds
to the star-formation rate in Figure 4; the dashed curve corresponds
to the star-formation rate derived by [24].  Figure 5 shows that
$N_*(z)$ peaks sharply at $z \approx 2$ and then drops off fairly
rapidly at higher $z$, with a tail that extends out to $z \approx 12$. 
The rapid rise in $N_*(z)$ out to $z \approx 2$ is due to the rapidly
increasing volume of space.  The rapid decline beyond $z \approx 2$ is
due almost completely to the ``edge'' in the spatial distribution
produced by the cosmology.  In essence, the sharp peak in $N_*(z)$ at
$z \approx 2$ reflects the fact that the SFR we have taken is fairly
broad in $z$, and consequently, the behavior of $N_*(z)$ is dominated
by the behavior of the co-moving volume $dV(z)/dz$; i.e., the shape of
$N_*(z)$ is due almost entirely to cosmology.  The right panel in
Figure 5 shows the cumulative distribution $N_*(>z)$ of the number of
stars expected as a function of redshift $z$.  The solid and dashed
curves have the same meaning as in the upper panel.  Figure 5 shows
that for the particular SFR we have assumed, $\approx 40\%$ of all
stars (and therefore of all GRBs) have redshifts $z > 5$.

\section{Conclusions}

If the long GRBs are indeed produced by the collapse of massive stars,
one expects GRBs to occur out to $z \approx 15-20$, redshifts that are
far larger than those expected for the most distant QSOs.  We have
shown that both GRBs and their afterglows are easily detected out to
these VHRs.  GRBs can therefore give us information about the
star-formation history of the universe, including the earliest
generations of stars.  The absorption-line systems and the Ly$\alpha$
forest visible in the spectra of GRB afterglows can be used to trace
the evolution of metallicity in the universe, and to probe the
large-scale structure of the universe at VHRs.  Finally, measurement of
the Ly$\alpha$ break in the spectra of GRB afterglows can be used to
constrain, or possibly measure, the epoch at which re-ionization of the
universe occurred.

\end{document}